\documentclass[useAMS,usenatbib]{emulateapj}

\usepackage{graphicx}

\newcommand{\msun}{${\rm M_{\sun}}$}

\def\ltsima{$\; \buildrel < \over \sim \;$}
\def\simlt{\lower.5ex\hbox{\ltsima}}
\def\gtsima{$\; \buildrel > \over \sim \;$}
\def\simgt{\lower.5ex\hbox{\gtsima}}
\def\kms{{\rm\,km\,s^{-1}}}

\def\kpc{{\rm\,kpc}}

\def\msun{{\rm\,M_\odot}}

\def\pc{{\rm\,pc}}

\newcommand{\fmmm}[1]{\mbox{$#1$}}
\newcommand{\scnd}{\mbox{\fmmm{''}\hskip-0.3em .}}
\newcommand{\scnp}{\mbox{\fmmm{''}}}
\newcommand{\mcnd}{\mbox{\fmmm{'}\hskip-0.3em .}}


\def\s{\ifmmode \widetilde \else \~\fi}
\def\={\overline}

\def\spose#1{\hbox to 0pt{#1\hss}}

\def\lta{\mathrel{\spose{\lower 3pt\hbox{$\mathchar"218$}}
     \raise 2.0pt\hbox{$\mathchar"13C$}}}
\def\gta{\mathrel{\spose{\lower 3pt\hbox{$\mathchar"218$}}
     \raise 2.0pt\hbox{$\mathchar"13E$}}}
\def\Dt{\spose{\raise 1.5ex\hbox{\hskip3pt$\mathchar"201$}}}    
\def\dt{\spose{\raise 1.0ex\hbox{\hskip2pt$\mathchar"201$}}}    

\def\dotsfill{\leaders\hbox to 1em{\hss.\hss}\hfill}

\def\ltsima{$\; \buildrel < \over \sim \;$}
\def\gtsima{$\; \buildrel > \over \sim \;$}
\def\lsim{\lower.5ex\hbox{\ltsima}}
\def\gsim{\lower.5ex\hbox{\gtsima}}
\def\lapp{\ifmmode\stackrel{<}{_{\sim}}\else$\stackrel{<}{_{\sim}}$\fi}
\def\gapp{\ifmmode\stackrel{>}{_{\sim}}\else$\stackrel{<}{_{\sim}}$\fi}

\slugcomment{Submitted to The Astrophysical Journal Letters}

\shorttitle{A central black hole in Sagittarius/M54}
\shortauthors{Ibata et al.}

\begin{document}

\title{DENSITY AND KINEMATIC CUSPS IN M54 AT THE HEART OF THE
SAGITTARIUS DWARF GALAXY: EVIDENCE FOR A $10^4
M_\odot$ BLACK HOLE?}

\author{R. Ibata\altaffilmark{1}, M. Bellazzini\altaffilmark{2}, 
S.C. Chapman\altaffilmark{3}, E. Dalessandro\altaffilmark{4}, 
F. Ferraro\altaffilmark{4}, M. Irwin\altaffilmark{3}, 
B. Lanzoni\altaffilmark{4}, G.F. Lewis\altaffilmark{5}, 
A.D. Mackey\altaffilmark{6}, P. Miocchi\altaffilmark{4}, 
A. Varghese\altaffilmark{1}}

\altaffiltext{1}{Observatoire Astronomique, Universit\'e de Strasbourg, CNRS, 11, rue de l'Universit\'e, F-67000 Strasbourg, France; ibata@astro.u-strasbg.fr}

\altaffiltext{2}{INAF - Osservatorio Astronomico di Bologna, via Ranzani 1, 40127, Bologna, Italy}

\altaffiltext{3}{Institute of Astronomy, Madingley Road, Cambridge CB3 0HA, UK}

\altaffiltext{4}{Dipartimento di Astronomia, Universita`degli Studi di Bologna, via Ranzani 1, I-40127 Bologna, Italy}

\altaffiltext{5}{Institute of Astronomy, School of Physics, A29 University of Sydney, NSW 2006, Australia}

\altaffiltext{6}{Institute for Astronomy, University of Edinburgh, Royal Observatory, Blackford Hill, Edinburgh EH9 3HJ, UK}

\begin{abstract}
We report the detection of a stellar density cusp and a velocity dispersion increase in the center of the globular cluster M54, located at the center of the Sagittarius dwarf galaxy (Sgr). The central line of sight velocity dispersion is $20.2\pm0.7\kms$, decreasing to $16.4\pm0.4\kms$ at $2\scnd5$ ($0.3\pc$). Modeling the kinematics and surface density profiles as the sum of a King model and a point-mass yields a black hole (BH) mass of $\sim 9400\msun$. However, the observations can alternatively be explained if the cusp stars possess moderate radial anisotropy. A Jeans analysis of the Sgr nucleus reveals a strong tangential anisotropy, probably a relic from the formation of the system.
\end{abstract}


\section{Introduction}
The relationships between the mass of central Supermassive Black Holes (SMBHs) and the physical properties of their host galaxies suggest a close connection between SMBH and galaxy formation and evolution \citep{Kormendy:1995p12081, Magorrian:1998p12173, Gebhardt:2000p12086, Ferrarese:2000p12098, Graham:2001p12161}. Determining the range of validity of these scaling relations and where they break down will likely lead to a deeper understanding of the link between BHs and their hosts. While numerous observational studies have demonstrated the presence of SMBHs in massive galaxies, there is only tantalizing evidence that the correlations also hold in low-mass systems, possibly down to globular cluster (GC) scales, where BHs with masses of a few $10^2-10^4 \msun$ (commonly named ``intermediate-mass BHs", IMBHs) have been suggested to reside (see references in \citealt{Lanzoni:2007p11564} and \citealt{Noyola:2008p12078}, hereafter N08). For low-mass galaxies the properties of the central BH do not appear to correlate well with host mass $M_{gal}$ (see, e.g., references in \citealt[][hereafter F06]{Ferrarese:2006p12063}). However, F06 have shown that a tight correlation exists with the so-called Central Massive Object (CMO), the latter being either a BH or a compact stellar nucleus: $M_{\rm CMO}\simeq 0.3\% M_{gal}$. 

Within this context the system comprising Sgr \citep{Ibata:1994p417} and the GC M54 presents an excellent prototype to study these issues on different mass scales. Sgr is a disrupting dwarf spheroidal, with a constant surface brightness core of about 1.7\kpc. At its center lies $(i)$ a stellar nucleus (hereafter Sgr,N) made of typical Sgr metal-rich stars with a scale-length $r_c = 0.4$ pc, and $(ii)$ M54, a very bright metal-poor GC ($r_c = 0.9$ pc, $M =2 \times 10^6 \msun$; \citealt{Pryor:1993p11793}).  As demonstrated in \citet[][hereafter  B08]{Bellazzini:2008p8100}, these two sub-systems are coincident in position and velocity, but exhibit different velocity dispersion ($\sigma$) profiles, thus suggesting that M54 was born in the halo of Sgr and was driven to coincide with Sgr,N by orbital decay via dynamical friction. Sgr is an excellent candidate for a system that would appear to have a compact stellar nucleus if it were located at the distance of the Virgo cluster. Indeed, the CMO of Sgr appears to be a combination of the central cluster M54 and Sgr,N, and the F06 relation suggests a mass for Sgr of $10^9\msun$, in good agreement with pre-disruption estimates \citep{Ibata:1998p389}. Here we push the investigation of the validity of the F06 relation to even smaller scales, addressing the possibility that M54 also hosts a CMO (an IMBH) in its center.  

\begin{figure*}
\includegraphics[angle=270, width=\hsize]{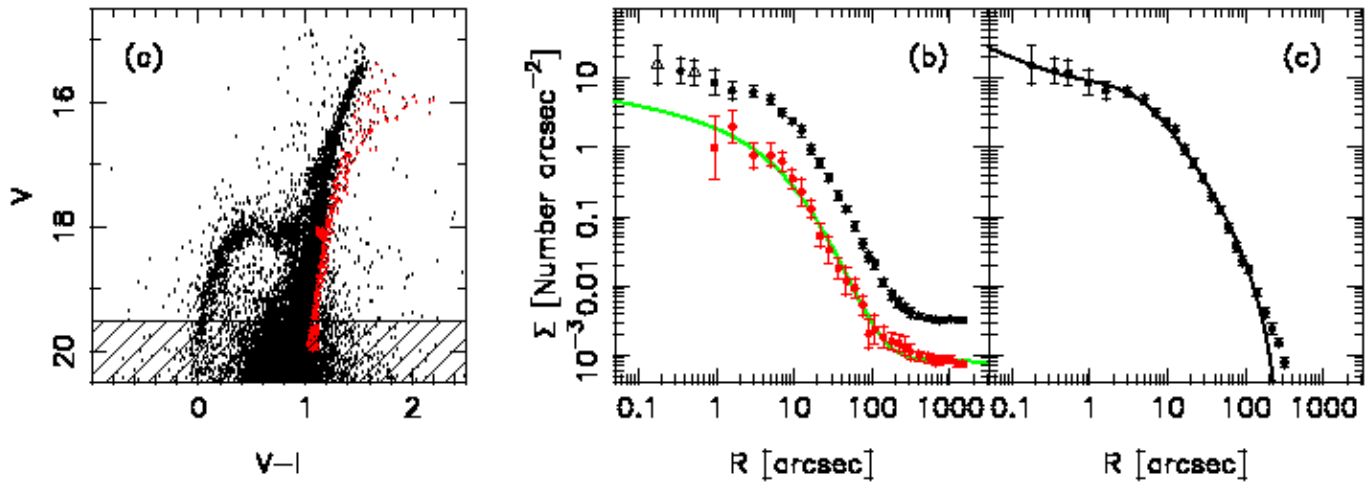}
\caption{(a) ACS color-magnitude diagram (CMD). Sources are classified into likely M54 (heavy black), Sgr (red), and foreground (light black), according to CMD location. (b) Corresponding star counts profiles for M54 (black) and Sgr (red), with respect to the M54 density center and as a  function of (projected) radius $R$. Towards the centre of M54 we detect a sharp rise in the stellar surface density, at odds with a flat distribution expected from a King model. The green line shows a Hernquist model fit to the Sgr profile (with background given by the King model of \citealt{Majewski:2003p4285}, also displayed in Fig.~2 of B08). The background-subtracted profile M54 is presented in (c), together with a composite model comprising a King model and a central IMBH.}
\end{figure*}

\section{Observations and Analysis}
\label{sec:data}

\subsection{Density Center and Profile}
To measure accurately the density profile and center of M54, we re-analysed the high-resolution dataset of B08, derived from Advanced Camera for Surveys (ACS) images taken in the $F606W$ and $F814W$ filters, with individual exposures of $t_{\rm exp}=30$\,s.  To improve the photometric and astrometric accuracy, the innermost $5\scnp$ were re-measured using ROMAFOT, a package specifically developed for crowded fields \citep{Buonanno:1989p11645}.  The catalog obtained by \citet[][hereafter M02]{Monaco:2002p11659} was used outside of the ACS field, while stars in common between the catalogs were used to calibrate the ACS photometry and astrometry ($0\scnd3$ uncertainty in absolute position).

The density center was measured using the \citet{Ferraro:2003p11665} procedure: from a first guess for the cluster center, we computed the mean of the positions of all stars brighter than a given limiting magnitude and lying within a circle of $\sim8-10\arcsec$ radius; this value was then used as the new centre for the search-circle, and the procedure was iterated until convergence. In order to limit contamination we adopted the color-magnitude selection criteria shown in Fig.~1a.  To avoid biases and spurious effects, we considered two samples with two different limiting magnitudes ($V<19$ and $V<19.5$). The values thus obtained agree within $0.2\arcsec$ and we adopted their mean ($\alpha = 18^{\rm h} 55^{\rm m} 3.345^{\rm s}$, $\delta =-30^{\circ} 28\arcmin 47\scnd1$) as the best estimate (cross-hairs in Fig.~2 mark this position). Note that this new determination is substantially different ($\Delta\alpha \sim 0\farcs7$ , $\Delta\delta \sim -5\arcsec$) from the \citet{Harris:1996p11754} center.

The projected density profile of M54 was measured by counting stars down to $V=19.5$ in 30 concentric annuli around this new center. Each annulus was divided into an adequate number of sub-sectors; the resulting density is the average of the corresponding sub-sector densities, and the quoted uncertainty is their rms dispersion. The star-counts profile is well matched by the surface brightness profile (which is unaffected by incompleteness); we therefore decided not to apply an incompleteness correction, avoiding an additional source of uncertainty.

The observed density profile of M54 is shown in Fig.~1b (black dots), where we have also plotted as triangles the values obtained by splitting the first bin ($r<0\scnd7$) in two portions (note that only 6 stars are found at $r<0\scnd35$).  If the innermost ($r\lsim1\arcsec$) points are excluded, the density profile is well-fit by an isotropic, single-mass King model with core radius $r_c\simeq 6\arcsec$ and concentration $c\simeq 1.8$. However, in the inner $\sim 1\arcsec$ there is an indication of a power-law deviation from a flat core behavior: the profile rises as
$\Sigma\propto R^{-\alpha}$, with $\alpha\simeq 0.3$. These values are consistent with those expected in the presence of a central IMBH \citep[see][]{Baumgardt:2005p11093,Miocchi:2007p11575}. Indeed, the observed profile is fit well by a King+BH model built as described in \citet{Miocchi:2007p11575}, with BH to total cluster mass ratio $M_{BH}/M = 0.0047$ (i.e., $M_{BH} \sim 9400 M_\odot$).

\begin{figure}
\includegraphics[angle=270, width=\hsize]{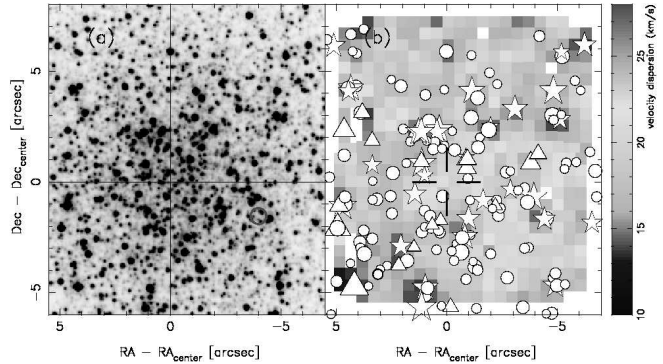}
\caption{(a) Center of the ACS I-band image. The density center is marked with cross-hairs, while the pixel with highest $\sigma$ is marked with a red circle. (b) Map of $\sigma$ over the same region, derived from the 2 deep ARGUS fields. The circles and ``star'' markers indicate the positions, respectively, of M54 and Sgr stars that are brighter than I=16 (triangles mark foreground sources). Bright stars tend to lower the measured $\sigma$ in the pixels whose light they dominate. Although the density center (0,0) has a high $\sigma$, a large region with coherently high dispersion is detected to the SW.}
\end{figure}

\subsection{Spectroscopy}
The fiber-fed FLAMES spectrograph at the Very Large Telescope (VLT) was used to follow up these findings. Two observing modes were employed: the ``MEDUSA'' and ``ARGUS" configurations, with the high-resolution HR21 mode, covering the \ion{Ca}{2} triplet. With MEDUSA, 130 fibers can be allocated over a 25\arcmin\ diameter field, while ``ARGUS'' covers a rectangular $22\times14$ pixel field ($11\scnd4 \times 7\scnd7$). Some observations from July and August 2005 were downloaded from the ESO archive, though most of the data were obtained on June 2--5 2008. All the data were reduced in a homogenous manner using the ESO pipeline ``Gasgano" software. A total of 944 isolated stars were observed with MEDUSA at radii between $0\mcnd5$ and $12\mcnd5$, while the entire inner region of the system out to a radius of approximately $0\mcnd6$ was covered with 51 overlapping ARGUS tiles, each of $t_{\rm exp}=300$\,s. A further 2 deep ARGUS fields ($t_{\rm exp}=2535$\,s) were secured on the very center of M54/Sgr under excellent seeing conditions ($\sim 0\scnd6$). The astrometry of the ARGUS data was refined by cross-correlating against the I-band ACS image.

\begin{figure}
\includegraphics[angle=0, bb= 30 15  540  540, clip, width=\hsize]{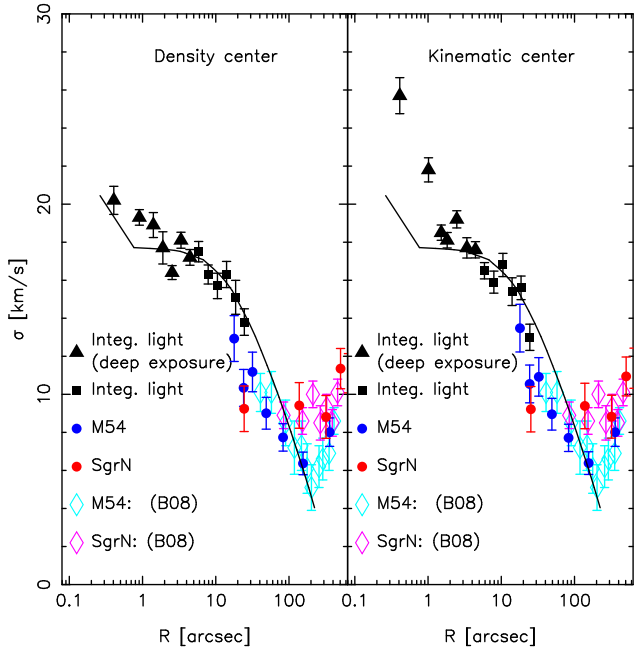}
\caption{The $\sigma$ profile measured from ARGUS spectra co-added in annuli (black triangles and squares) rises rapidly towards the cluster center. Beyond $\sim 20\scnp$, the MEDUSA and ARGUS spectra can be used to measure the velocities of individual stars, uncontaminated by neighbors. The blue and red points show the corresponding $\sigma$ profiles for the M54 and Sgr samples, respectively. For comparison, we show the earlier B08 results. The line shows the kinematic behavior of the King+BH model shown previously in Fig.~1c.}
\end{figure}

All the individual MEDUSA and ARGUS fibers were sky-subtracted and cross-correlated against an artificial template as in \citet{Ibata:2005p216}. Velocity uncertainties, estimated from the r.m.s. scatter in Gaussian fits to each \ion{Ca}{2} line independently, are typically below $1\kms$ for bright stars.

Towards the center of M54 our limited spatial resolution causes the superposition of many stars under a single fiber. The dispersion in velocities of these stars results in a broadening of the observed spectral lines, which can be used to measure the stellar kinematics. We used the Penalized Pixel-Fitting (pPXF) method of \citet{Cappellari:2004p11794} to perform the line of sight velocity distribution deconvolution. Bright isolated stars, observed with the same instrument, were selected to serve as spectral templates.

\subsection{Velocity dispersion profile}

Fig.~2b shows the $\sigma$ map derived from the two deep ARGUS fields using the pPXF algorithm. The map must be interpreted with caution, since we are in a regime where some pixels are dominated by one or a few bright stars. Clearly, where a single star dominates, $\sigma$ will tend to be abnormally low; furthermore, spuriously large $\sigma$ values can be measured in pixels containing only a few stars with velocities on opposite tails of the velocity distribution. However, the velocity distribution will tend towards more reliable values when spectra are co-added over several spatial pixels.

As is apparent from Fig.~2b, there is a noticeable correlation between bright stars and those pixels with low $\sigma$ values. However, one also perceives a radial decrease in $\sigma$. Surprisingly, the highest $\sigma$ peak (at $[-4\scnp,-1\scnd5]$) does not lie at the density center of the system. Although this peak may be an artifact of a chance velocity configuration of the 4 bright stars that surround it, it nevertheless lies in the middle of a $\sim 3\scnp$-wide high $\sigma$ region, so the reality of the peak cannot easily be dismissed. Inspection of the ACS image (Fig.~2a) reveals that this ``kinematic center'' lies outside any obvious density peak. We verified that this does not correspond to the center of the metal-rich Sgr stars.

\begin{figure}
\includegraphics[angle=0, bb= 30 20 540 310, clip,
  width=\hsize]{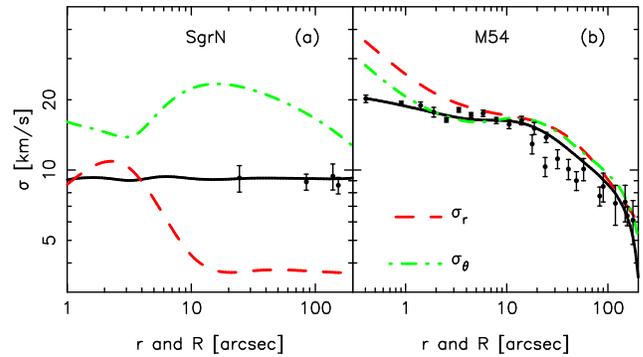}
\caption{(a) An MCMC scheme was used to find the most likely profiles for the radial (red) and tangential (green) components of $\sigma$ (as a function of radius $r$) of the Sgr,N population that give rise to a projected $\sigma$ that is flat as a function of projected radius $R$. (b) The components of $\sigma$ required of the M54 population if no BH is present. (Data points are reproduced from Fig.~3).}
\end{figure}

Bearing this ambiguity in mind, we present the $\sigma$ profile over the entire radial range of M54 in Fig.~3, with panels (a) and (b) obtained with respect to the density and kinematic centers, respectively. Circles and lozenges mark the $\sigma$ values derived from isolated stars observed in single fibers; these values and their uncertainties were calculated using a maximum likelihood method (see B08). Squares and triangles are derived from ARGUS spectra co-added in annuli, and measured with the pPXF software; the corresponding uncertainties were estimated by ``bootstrap-resampling'' \citep{Press:1992p11802} the spectra, and re-measuring $\sigma$ in 100 random simulations at each radial bin. ARGUS pixels within $0\scnd7$ of a star of magnitude $I=15$ were rejected. The line shows the prediction of the King+BH model presented in Fig.~1, and gives a close representation of the measured $\sigma$ profile.

The $\sigma$ profile of the Sgr population is also shown in Fig.~3. The isothermal behavior of this population seen by B08 at larger radii clearly continues inwards to $R\sim 20\scnp$.

\subsection{Anisotropy of Sgr,N}
The data presented above allow us to address a conundrum raised in B08: how is it possible for M54 (with a rising $\sigma$ profile towards the centre) to co-exist in equilibrium with the Sgr,N population, that exhibits an isothermal projected profile, almost constant at $\sigma=9.2\kms$?  To answer this question, we turn to the Jeans equation for a spherically-symmetric system (\citealt{Binney:1987p11821}, Eqn 4-55):
\begin{equation}
{{G M(r)}\over{r }} = - \sigma_r^2 \Bigg[ {{d \ln \rho}\over{d \ln
        r}} + {{d \ln \sigma_r^2}\over{d \ln r}} + 2\Big(1 -
    {{\sigma_\theta^2}\over{\sigma_r^2}} \Big) \Bigg] \, ,
\end{equation}
where $M(r)$ is the cumulative mass inside radius $r$, $\rho$ is the density, and $\sigma_r$ and $\sigma_\theta$ are the radial and tangential velocity dispersions, respectively.  Interestingly, given the Sgr density profile shown in Fig.~1b, if we assume that the Sgr,N population is isotropic, the total mass profile resulting from the Jeans equation in the radial range 1--30\scnp\ turns out to be substantially lower (by a factor up to $\sim 10$) than that of the King model (Figs.~1c and 3a) that best fits the M54 population alone. Since this is physically absurd, we conclude that the Sgr,N population cannot have isotropic orbits.

We can now turn this argument around, to ask what level of anisotropy is consistent with the M54 mass model, since the cluster likely represents the dominant component in the center of the M54/Sgr system at $R \lsim 100\scnp$ (Sgr stars and accompanying dark matter should be relatively unimportant in this region). Hence, by assuming the M54 King model for $M(r)$, and the Hernquist model fit to the Sgr population for $\rho$, $\sigma_r$ and $\sigma_\theta$ remain as the only unknowns in the Jeans equation. However, $\sigma_r$ and $\sigma_\theta$ are constrained by the observed (projected) dispersion $\sigma$. A trial profile for $\sigma_r(r)$, defined by a bi-cubic spline at 5 logarithmically-spaced points (red line in Fig.~4), was improved upon iteratively using a Markov-Chain Monte-Carlo (MCMC) scheme with ``parallel-tempering'' chains (see \citealt{Gregory:2005p11863}); the set target was to produce a flat projected $\sigma(R)=9.2\kms$.

Fig.~4a shows that the nested system can be constructed, but it requires a high degree of anisotropy for Sgr,N, with orbits being exceedingly tangential between 20--200\scnp (interior to 20\scnp\ the fit is based only an extrapolation of the $\sigma$ profile of Sgr,N). If M54 decayed into the center of Sgr, as suggested by B08, its orbital energy must have been transferred to stars previously present in that region. It is tempting to assume that that this would manifest itself as a high $\sigma_\theta$, as inferred.

\section{Discussion and Conclusions}

The star-counts profile of M54 (Fig.~1c) shows a sharp rise in the inner 1\scnp, a behavior that is also mirrored in the $\sigma$ profile (Fig.~3a). Although the rise up to $\sigma \sim 18\kms$ at $R=2\scnp$ is a robust result, the kinematic data are not as clear-cut interior to this radius. It is possible that the stochastic distribution of stars in both projected position and line of sight velocity could give rise to spuriously high $\sigma$ values. Indeed, we judge the ``kinematic peak'' to be such an artifact, since it appears to have no density counterpart. However, the central rise in $\sigma$ around the density center is in agreement with expectations from the stellar density cusp, and since this region is relatively free of very bright stars that could bias the $\sigma$ measurements, we judge that the measured rise in $\sigma$ is likely real. Further support for the reality of the M54 $\sigma$ peak comes from a comparison to $\omega$ Centauri. Recently, N08 analyzed a data-set comprising ACS photometry and Gemini integral-field spectroscopy, finding evidence for both a power-law increase in the central stellar profile, as well as a marked rise to $\sigma=23\kms$ in the center, which they interpret as providing evidence for a $4\times 10^4 \msun$ BH. A reanalysis by \citet{Miocchi09} finds $M_{BH}/M \sim 0.0057 \pm 0.001$, which (depending on the adopted cluster mass) brings down $M_{BH}$ to 1/7 to 1/2 of the N08 value, consistent with the \citet{Marel:2009p12926} $18000\msun$ upper limit. The similarity to the results presented above is striking, all the more as it suspected that $\omega$ Cen is the stripped core of a galaxy that was once similar to Sgr (\citealt{Bekki:2003p12175}, B08). While it could be argued that the central dispersion increase in one of these systems could be due to chance stellar alignments, it would require a conspiracy for both to be affected in the same way (especially since the discreteness of the underlying stellar distribution should in general {\it lower} $\sigma$). 

What is the nature of this central structure? Is it due to a cold dark matter (CDM) cusp, a population of massive stellar remnants, peculiar stellar kinematics, or indeed an IMBH?  For any reasonable total mass and concentration parameters of Sgr, a \citet{Navarro:1997p12185} CDM cusp would possess only $\sim 10^3\msun$ within the inner 10\scnp, and so cannot account for the $\sigma$ rise. As discussed in \citet{Noyola:2008p12078}, neither can massive stellar remnants be accommodated, because of the implausibly high central concentration required if the cluster has not undergone core-collapse (like M54 and $\omega$ Cen).  

In order to explore the alternative that orbital anisotropy is the cause of the observed high central $\sigma$, we performed a Jeans analysis similar to that described above. We assumed the stellar density profile shown in Fig.~1c and the corresponding mass profile (leaving total mass as a free parameter).  As before, we iterate on an initial guess for $\sigma_r(r)$ (taking as a prior that M54 should be isotropic beyond 10\scnp), to find the projected $\sigma(R)$ that is consistent with the kinematic observations of M54. The results are shown in Fig.~4b: significant, but by no means extreme, radial anisotropy ($\sigma_r/\sigma_\theta=1.25$ at 1\scnp) is required in the central 2\scnp\ to reproduce the rising $\sigma$. Rejecting a single strongly outlying data-point (at $24\scnp$), the reduced-$\chi^2$ of this model ($\chi^2=2.7$) is marginally worse than that of the King+BH model of Fig.~3a ($\chi^2=2.5$). While very short relaxation times in the centre of a cluster are commonly believed to lead to isotropic stellar orbits, this might not hold if there is a strong density gradient (i.e. if there is a cusp). The IMBH solution appears the only viable alternative, however, if the orbits must be isotropic.

As discussed in \citet{Maccarone:2008p12070}, stellar winds should provide gas to the central BH and make it emit as a radio or X-ray source.  Due to the possible contribution of Sgr gas, such emission might be more significant in the case of the M54/Sgr system, compared to normal GCs. However, the expected radio and X-ray fluxes are still highly uncertain.  While no suitable radio observations are available for M54, \citet{Ramsay:2006p11898, Ramsay:2006p11902} have studied its X-ray emission using the Chandra Observatory. We retrieved their data from the archive and compared the position of the X-ray bright sources to the ACS image. Interestingly, with the default astrometric solution provided by the Chandra pipeline, their object \#2 lies within 1\scnp\ of the density center of M54. Since the Chandra absolute astrometric accuracy is 0\scnd6\footnote{\tt http://cxc.harvard.edu/cal/ASPECT/celmon/}, while the systematics in the absolute position of the ACS are $\sim 0\scnp3$, it is possible that this source may be associated with the stellar cusp we have identified. From their luminosities and colors, \citet{Ramsay:2006p11898} suggest that the 7 sources they detected in M54 are likely cataclysmic variables or low-mass X-ray binaries. However, their object \#2 stands out as being the only irregularly-shaped source in this small sample, and with an X-ray luminosity of $L_X= 0.72\times 10^{33} {\rm erg s^{-1}}$ is bright enough to lie at the lower limit of the \citet{Maccarone:2008p12070} predictions. 

The data discussed above do not allow us to clearly state whether the putative IMBH belongs to M54 or to Sgr. However, we note that the estimated BH mass is consistent with the (extrapolation of the) F06 relation only if its host system is M54 since $M_{BH}/M_{M54}\sim 5\%$, while a BH 1000 times more massive would be needed for a similar mass ratio with respect to Sgr.  We therefore may be in the presence of a very interesting system composed of a dwarf galaxy hosting a prominent stellar nucleus, itself hosting a central IMBH, and both following the F06 relation accurately-verified at higher mass regimes.  

If the presence of an IMBH can be confirmed, it would have far-reaching implications not only for the formation of SMBHs and galaxies, but also for a deeper comprehension of the unclear nucleation process. However, as we have shown, the observations can be equally well explained by a centrally-concentrated stellar population with radially-anisotropic orbits. It will therefore be very useful to explore with N-body simulations whether orbital anisotropy can be maintained in such a dense system, and to obtain higher-spatial resolution spectroscopy to probe closer to the central mass.


\end{document}